# Optimizing Broadband Terahertz Modulation with Hybrid Graphene/Metasurface Structures

S.-F. Shi,[1,2,*,¶] B. Zeng,[1,¶] H.-L. Han,[1,¶] X. Hong, H.-Z. Tsai,[1] H. S. Jung,[1] A. Zettl,[1,2,3] M. F. Crommie,[1,2,3] F. Wang[1,2,3,*]

[1] Department of Physics, University of California at Berkeley, Berkeley, 94720

[2] Material Science Division, Lawrence Berkeley National Laboratory, Berkeley, 94720

[3] Kavli Institute at University of California at Berkeley, Berkeley, 94720

[¶]These authors contribute equally to this work

[*]Corresponding author's email address: Su-Fei Shi (sufeishi@berkeley.edu),

Feng Wang (fengwang76@berkeley.edu)

CORRESPONDING AUTHOR FOOTNOTE:
Email: sufeishi@berkeley.edu, fengwang76@berkeley.edu

Telephone number: (510) 643-3275
Fax number: (510) 486-6054

**Abstract:** We demonstrate efficient terahertz (THz) modulation by coupling graphene strongly with a broadband THz metasurface device. This THz metasurface, made of periodic gold slit arrays, shows near unity broadband transmission, which arises from coherent radiation of the enhanced local-field in the slits. Utilizing graphene as an active load with tunable conductivity, we can significantly modify the local-field enhancement and strongly modulate the THz wave transmission. This hybrid device also provides a new platform for future nonlinear THz spectroscopy study of graphene.

Terahertz (THz) wave is widely referred to as the "last frontier"[1] in the electromagnetic spectrum, with frequency ranging from 0.3 to 10 THz (1 THz = $10^{12}$ Hz) and photon energy on the order of meV. For the interest of fundamental research, it facilitates spectroscopic studies[2] of low energy excitations in molecules[1,2] and superconductors[3,4]. For the interest of applications, it is viewed as the foundation to next generation of noninvasive imaging[5–7] and wireless communication[8]. Although THz technology is still immature compared to its electronic and optical counterparts, we have seen significant progress in THz generation[9–11] and detection[12,13] over the last decade which leads to the surging demand for active THz wave devices[14–16]. Among them, an electrically controlled THz modulator is a key component for real time manipulation of THz waves. Graphene, a single layer of carbon atoms arranged in honeycomb structures, is a promising candidate for this application because of its large and tunable absorption in THz regime. Recently, it has been shown that graphene in a field effect transistor (FET) configuration provides broadband THz power modulation of ~ 20%[17], a significant improvement over previous semiconductor modulators. This broadband modulation is advantageous given the limited THz sources available. However, its modulation depth is still constrained by the finite conductivity achievable for a single atomic layer.

Plasmonic structures[18,19], on the other hand, exhibit extraordinary light transmission and greatly enhanced local electrical field. By modifying this enhanced field, it is possible to achieve large modulation of transmitted waves. Atomically thin and flexible, graphene can efficiently couple to locally enhanced fields, known as "hot spots". Consequently we can achieve large modulation with a graphene/plasmonic-structure hybrid device using graphene as an active load[20,21]. However, there remain two major challenges to implement graphene/plasmonic-structure hybrid device for optimized THz modulation. First of all, it is desirable to preserve the broadband THz modulation of bare graphene for the new hybrid device[17,22]. Secondly, an efficient coupling between graphene and the plasmonic structure is necessary for optimized modulation. For example, graphene conductivity is attributed to the delocalized π-electrons[23,24] so that the enhanced local field needs to be engineered "in-plane" to maximize the coupling efficiency. Recently, semiconductors[25] or graphene[26,27] coupled to resonant THz metamaterials have shown improved modulation depth but they only work for a narrow frequency window. Meanwhile, researchers have attempted to improve the THz modulation by coupling graphene to broadband metamaterials[28], but the modulation depth is still limited and the study on optimized coupling is lacking.

In this paper, we address these challenges by optimizing the coupling between graphene and a new type of metasurfaces (two dimensional metamaterial) made of a periodic array of gold slits. Such THz metasurfaces were proposed theoretically[29] and show close to unity THz transmission over a broad frequency range. The large transmission arises from the coherent radiation of greatly enhanced in-plane local field in the slits, which efficiently couples to graphene. Adopting the idea of an active load, we propose to modify the local-field by tuning the conductivity of graphene in this strong coupling regime and we show that THz modulation can be greatly enhanced by properly designing the slit arrays. We verify in both simulation and experiment that an enhanced broadband THz modulation is achieved in the optimized graphene/metasurface hybrid device using electrostatic gating. We also show that a strong coupling with the metasurface gives rise to an enhanced absorption in graphene at the charge neutral point (CNP).

This hybrid graphene-metasurface device therefore can be useful for nonlinear THz investigations of graphene considering the limited intense THz sources available.

The metasurface of periodic gold slit arrays with a period of $P$ and a slit width of $w$ is shown schematically in Figure 1a. The local field in the slit is strongly enhanced due to capacitive charging of the narrow slit when the normal incident light is polarized perpendicular to the slit[29,30]. Coherent radiation from the oscillating local field interferes constructively and leads to a pronounced far field transmission. When the metasurface is suspended in air, the far field transmission and local field enhancement is related in a simple form as follows (Supplementary Information):

$$T = \left(\frac{\eta w}{P}\right)^2 \quad (1)$$

where $T$ is the power transmission, $\eta$ is the average enhancement factor of the local-field in the slit defined as:

$$\eta = \frac{\int_0^w E dx}{w E_0} \quad (2)$$

In Eq. (2), $E_0$ is the electrical field of the incident light. For an optimized design with $\frac{w}{P} \ll 1$ and $\frac{P}{\lambda} \ll 1$, we have $\eta = \frac{P}{w}$, which directly leads to 100% transmission[31,32] for the metasurface despite a geometry ratio $\frac{w}{P}$ much less than 1. This near-unity transmission is a direct consequence of the greatly enhanced local electrical field in the slit, which is frequency independent and thus broadband.

From Eq. (1), a large modulation of the transmission can be achieved by tuning the graphene conductivity to modulate the local field enhancement $\eta$. The enhancement of transmission modulation in the graphene/metasurface hybrid device can be easily understood in the perturbation regime, where the decreased transmission $\Delta T$ results mainly from the change in graphene absorption $\Delta P_{abs}$, namely $\Delta T = -\Delta P_{abs}$. For bare graphene, $\Delta P_{abs} = AE_0^* \cdot J = A|E_0|^2 \Delta\sigma'$, where A is the graphene area and $\sigma'$ is the real part of the optical conductivity[33]. For the hybrid device, only graphene in the slit (with a geometry ratio $\gamma = w/P$) absorbs light. Considering the average field in the slit $\langle E_{slit} \rangle = \eta E_0$, we have $\Delta P_{abs} = \gamma A \times \langle E_{slit}^* \cdot J \rangle \approx \gamma A |E_0|^2 \Delta\sigma' \eta^2$. An optimized design with $\eta \approx P/w$ leads to

$$\Delta P_{abs} \approx A|E_0|^2 \Delta\sigma' \frac{P}{w} \quad (3)$$

It is clear that the absorption change (and the change of the transmission) is enhanced by a factor of $\frac{P}{w}$ compared to that of bare graphene. This large absorption enhancement can also be used for nonlinear THz spectroscopy studies of graphene where a strong light matter interaction is usually required.

To verify Eq. (1) in a general context, we simulate THz transmission of the slit arrays for various geometries using finite element methods (see Supplementary Information). For simplicity, we assume to have periodic gold slits suspended in air (Fig. 1a). The simulation result is shown in a two-dimensional plot with different $P/\lambda$ (x-axis) and $w/P$ (y-axis) values (Fig. 1c). It exhibits nearly 100% transmission over a broad range of frequencies even for slit arrays with small geometry ratio ($w/P \ll 1$). Figure 1d displays frequency-dependent transmission spectra for a few different geometries. With slit width fixed at 2 $\mu m$, the "crossover" frequency, defined as the excitation wavelength at which transmission of slit arrays falls to ~50%, shifts to longer wavelength as we increase the period from 20 $\mu m$ to 60 $\mu m$. For slits with a fixed period of 20 $\mu m$ (with various slit width of 2 $\mu m$, 4 $\mu m$ and 6 $\mu m$), the transmission remains near unity from 0 to 2 THz. In particular, for the 2-20 $\mu m$ design (slit width 2 $\mu m$ and period 20 $\mu m$) our simulation shows a greatly enhanced local field in the slit (Fig. 1b) with an average enhancement factor $\eta$ of 9.6 at 1 THz, consistent with the prediction ($P/w = 10$) of Eq. (1).

Experimentally we fabricate our periodic gold slit array devices with various geometries on silicon substrate with 1.8 $\mu m$ thermal oxide and we measure the transmission as a function of frequency in the window 0.4-2.0 THz. We observe near unity transmission (Fig. 2a) for the 4-20 $\mu m$ device (width 4 $\mu m$ and period 20 $\mu m$, dashed magenta line) and the 6-20 $\mu m$ device (dashed red line). The experimental data agree well with simulation results (solid traces in Fig. 2a), with substrate effect considered. For fixed slit width of 4 $\mu m$ or 6 $\mu m$ (Fig. 2a) we found that the "cross-over" shifts to longer wavelength as the period increased, as predicted by the theory[29,31]. Remarkably, THz transmission of the metasurface device is also strongly polarization dependent. The transmission for the 4-20 $\mu m$ device is at maximum when the polarization of light is perpendicular to the slit and at minimum when the polarization is parallel to the slit. The extinction ratio, defined as the magnitude ratio between two transmitted THz fields of different polarization ($|E_\perp|/|E_\parallel|$), is as high as ~1000. This is consistent with Eq. (1) since a large $\eta$ only occurs when the incident light induces charge oscillations across the slit.

Next, we measure THz modulation of the proposed graphene/metasurface hybrid device. We fabricate the hybrid device by transferring a single layer graphene grown by chemical vapor deposition (CVD)[34] method to a 2-20 $\mu m$ gold slit arrays on a SiO2/ Si substrate. The 2-20 $\mu m$ geometry, relatively easy to fabricate with standard photolithography, is chosen for both large THz transmission and local field enhancement $\eta$ to optimize the THz modulation. We define the source, drain and gate electrodes on graphene using a shadow mask and gate graphene electrostatically using ion-gel[35], as schematically shown in Figure 3a. During the THz transmission measurement, we vary the gate voltage to control the density of carriers in graphene and simultaneously monitor the resistance of the device by applying a small bias across the source-drain electrodes. This simultaneous electrical transport measurement helps to determine the charge neutral point (CNP) of graphene (~ 0.33 V in Fig. 3c). We observe that as we gate graphene from CNP (blue trace, Fig. 3b) to -1.75 V away from CNP (magenta trace, Fig. 3b), the transmission of THz wave decreases appreciably. Since the THz transmission spectrum of the device is relatively flat in the frequency window of our measurement[19,33,35], we can use the change of the field peak value to estimate the power transmission of the hybrid device, which is roughly ~84% when the graphene is at CNP and ~ 43% for graphene gated at -2 V away from the CNP. By monitoring the peak value change, we obtain the normalized transmission change as a function of the gate voltage (Fig. 3d). This change of transmission through the hybrid device (red dots in Fig. 3d) closely correlates to the graphene conductance that we measure simultaneously (Fig. 3c): the transmission is the highest when the graphene is charge neutral (least conductive) and the lowest when the graphene is most conductive.

This large modulation of the THz wave transmission is associated with the modification of the local-field in the slit by the graphene. Here, graphene acts as an active load to the slit arrays. Highly doped graphene provides a direct conducting channel that shorts the charge oscillations across the slit. This "shorting"[36] significantly suppresses the local field enhancement, reducing THz wave transmission according to Eq. (1). This is confirmed by our simulation in Figure 4b which shows that the field enhancement in the slit of the 2-20 $\mu m$ device is the highest when there is no graphene (0 $G_0$), and the average field enhancement factor significantly decreases when the graphene in the slit is gated from conductivity of 4 $G_0$ (minimum conductivity at CNP[37]) to 60 $G_0$. Quantitatively, our simulation shows that the average enhancement factor $\eta$ of the local-field in the slit for the case of no graphene is ~ 8.8 for an incident wave at 1 THz (see Supplementary Information). When the graphene is gated to 4 $G_0$, $\eta$ remains largely unchanged (the enhancement factor decreases slightly to ~8.0). However, when we dope the graphene to high conductivity of 60 $G_0$, we significantly reduce $\eta$ to ~3.3. This reduction of local field enhancement leads to our observed THz transmission modulation for this hybrid device. To quantitatively determine the improvement of THz modulation of the graphene/metasurface hybrid device over bare graphene, we plot the normalized THz power transmission (normalized to the transmission for charge neutral graphene) as a function of graphene conductivity at 1 THz. With the large scattering rate of CVD graphene, we can approximate the graphene conductivity at 1 THz to be the DC conductivity of graphene[33]. We measure the graphene DC conductance through two terminal transport measurements[38] with a small voltage bias, and extract the conductivity of the graphene sheet while taking into account the geometry ratio of the device and the contact resistance (Supplementary Information). We show in Figure 4c that the normalized THz transmission for bare graphene (red empty dots) and the hybrid graphene-metasurface (2-20

$\mu m$) structure agree well with simulation results (solid traces). It is obvious that at 20 $G_0$, which can be comfortably achieved by many solid state graphene devices, the graphene/2-20 $\mu m$-slit-arrays hybrid device has a power modulation depth (defined as $(T_{max} - T_{min})/T_{min}$) of 0.9, which is 9 times larger than that of bare graphene. At a higher graphene conductivity of 70 $G_0$, the hybrid device has a power modulation of 5.9, more than 20 times larger than that of bare graphene. Theoretically, performance of this hybrid THz modulator can be further advanced following our optimization guidelines discussed in Eq. (3). By using a metasurface with smaller slit width ($P = 2 \ \mu m, w = 0.1 \ \mu m$) the modulation depth (green solid trace in Fig. 4b) can be as high as 2.4 and 17 for graphene's conductivity of 20 $G_0$ and 70 $G_0$ respectively, showing an additional 3 times enhancement over the 2-20 $\mu m$ device.

In summary, we systematically investigate, both in theory and in experiment, the efficient coupling of a broadband THz metasurface to graphene. We demonstrate experimentally improved broadband THz wave modulation using an optimized design of the graphene/metasurface hybrid device. In addition, strong THz absorption and a large local field enhancement exist in nearly charge neutral graphene. Thus, this hybrid device can potentially provide a platform for strong light matter interactions in graphene and in nonlinear THz studies.

# Methods

**Device fabrication.** We fabricate the periodic gold slit arrays using standard photolithography followed by ebeam evaporation of 5 nm/80 nm Ti/Au and lift off. We use low doped silicon (resistivity 10-20 Ohm cm) with thermal oxide as the substrate for our THz transmission measurement. We grow single layer graphene on copper foil using the chemical vapor deposition (CVD) method. After etching copper away, we transfer the graphene on top of the gold slit arrays. We evaporate 5/50 nm Ti/Au through a shadow mask to define source, drain and gate electrodes. We use ion-gel as the gating dielectrics to control the doping of graphene.

**THz transmission measurement.** We generate THz waves using air-plasma method where 800 nm and 400 nm laser pulses (both with ~160 fs pulse duration and 1 kHz repetition rate) are simultaneously focused at the same spot in air. The generated THz radiation is collimated and focused onto our sample through a pair of parabolic mirrors. The transmitted THz wave is then re-collimated and focused onto a ZnTe crystal of 0.5 mm thickness, where the THz electrical field waveform is detected using electro-optic sampling. We use ion-gel gating to control the carrier concentration in graphene and the resistance of the graphene is monitored in-situ during THz transmission measurement.

# Acknowledgement

We thank Dr. Qin Zhou and Halleh Balch for help. Optical characterization of this work was mainly supported by Office of Basic Energy Science, Department of Energy under Contract No. DE-SC0003949 and DE-AC02-05CH11231 (Subwavelengths metamaterials). Graphene synthesis and photonic device fabrication were supported by the Office of Naval Research (Award N00014-13-1-0464). We also acknowledge the support from a David and Lucile Packard fellowship.

Supporting Information Available: [Details about the theory, simulation, sample preparation, and data analysis.] This material is available free of charge via the Internet at http://pubs.acs.org.

**Figure 1. Broadband terahertz (THz) metasurface of periodic gold slits.** (a) Schematic drawing of periodic gold slits suspended in air. (b) Simulated spatial distribution of electrical field enhancement of THz wave for a gold slit of width 2 $\mu m$ and period 20 $\mu m$. The THz wave is polarized along $x$ (perpendicular to the slit). (c) Color plot of power transmission as a function of $\frac{P}{\lambda}$ (period to wavelength ratio) and $\frac{w}{P}$ (slit width to period ratio). (d) Frequency dependent transmission for various geometries. The three solid lines correspond to the dashed line cuts in (c) and show large transmission over a broad range of frequency (black trace is for slit width 2 $\mu m$ and period 20 $\mu m$). Slits with width 2 $\mu m$ and period 40 $\mu m$ (red dashed trace) and 60 $\mu m$ (magenta dashed trace) show decrease of transmission at high frequency. All simulations are performed for gold slit devices which are 80 $nm$ thick and suspended in air.

**Figure 2. Experimental demonstration of the THz metasurface.** (a) THz power transmission as a function of frequency for various geometries. The solid dots and squares are experimental data and the dashed lines are results from simulations. (b) Experimental THz power transmission as a function of the angle between THz wave polarization and the slit orientation for a gold slit device of width 4 $\mu m$ and period 20 $\mu m$. All devices are supported on low doped silicon chip with 1.8 $\mu m$ thick SiO$_2$ and have gold thickness of 80 $nm$.

**Figure 3. THz modulation by the graphene/metasurface hybrid device.** (a) Schematic representation of the experiment setup for measurement as well as the hybrid device configuration. We fabricate the hybrid device by transferring a single layer graphene on top of a gold slit device of width 2 $\mu m$ and period 20 $\mu m$. The incoming THz wave is polarized

perpendicular to the slit orientation. (b) Transmitted THz waveforms for a reference sample (dashed black trace), the hybrid device with graphene at CNP (solid blue trace), and the hybrid device with graphene at gate voltage -1.75 V (solid magenta trace). (c) Resistance measured for this hybrid device shows CNP at the gate voltage of 0.33 V. The side view of a schematic representation of our ion-gel gated device is shown in the inset, where the blue line represents graphene.(d) Simultaneously measured THz field transmission of this hybrid device (normalized to the transmission when graphene is at CNP).

**Figure 4. THz modulation depth vs. graphene conductivity.** (a) Schematic drawing of the side view of the hybrid device. (b) Simulated field enhancement factor in the slit for the hybrid device when there is no graphene (0 $G_0$), graphene conductivity at 4 $G_0$, and graphene conductivity at 60 $G_0$, respectively. The slit device is of width 2 $\mu m$ and period 20 $\mu m$. (c) Normalized THz power transmission modulation as a function of graphene conductivity. The solid line traces are simulation results for bare graphene (red) and the hybrid device (blue). The empty dots are experimental data for bare graphene (red) and the hybrid device (blue, magenta).The conductivity of the hybrid device is determined from DC transport measurement (Supplementary Information).

Figure 1.

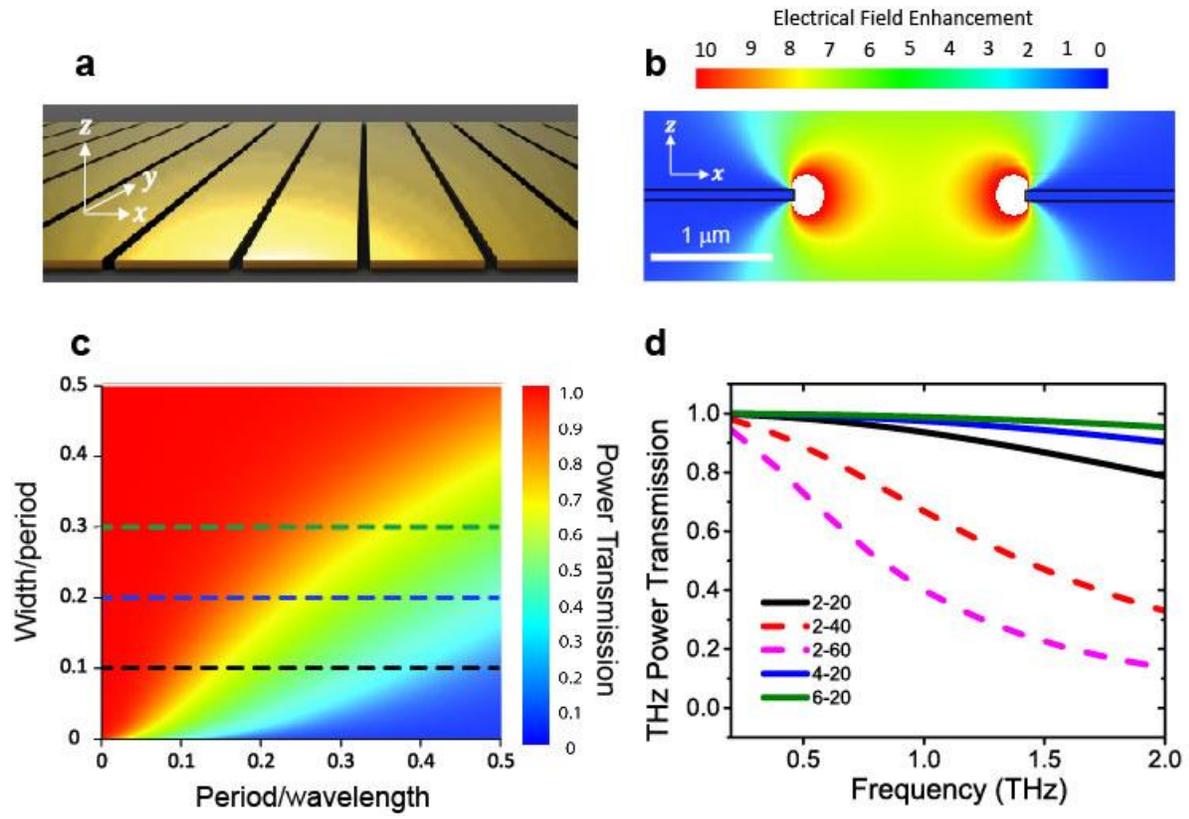

Figure 2.

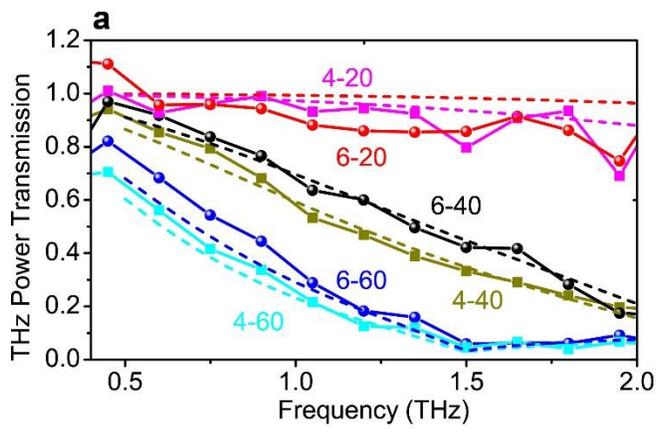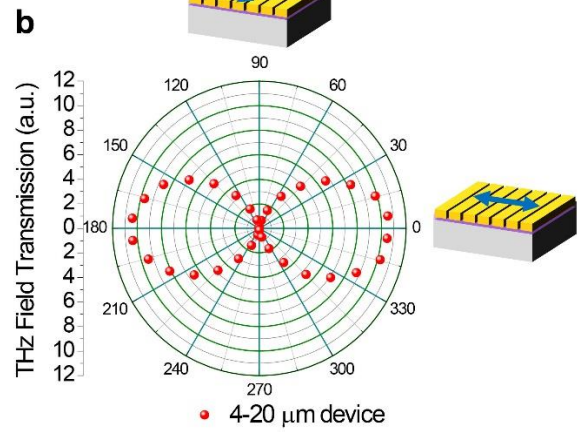

Figure 3.

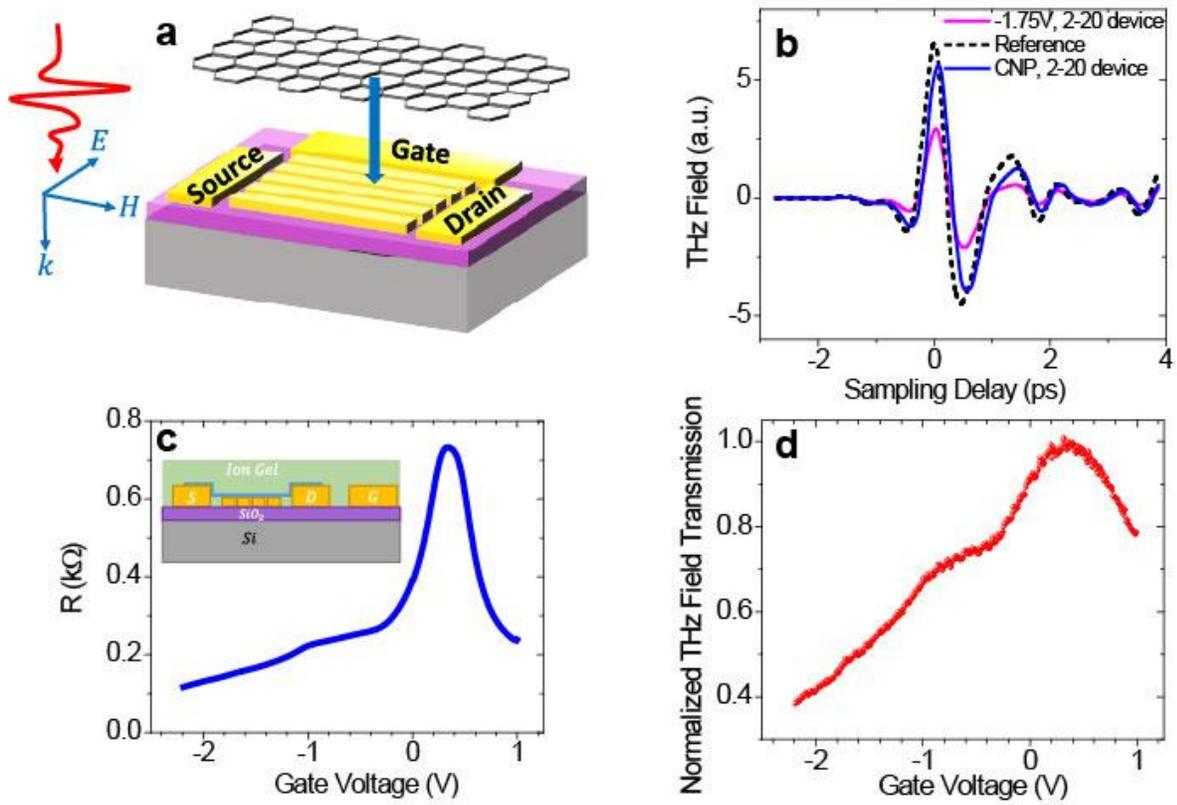

Figure 4.

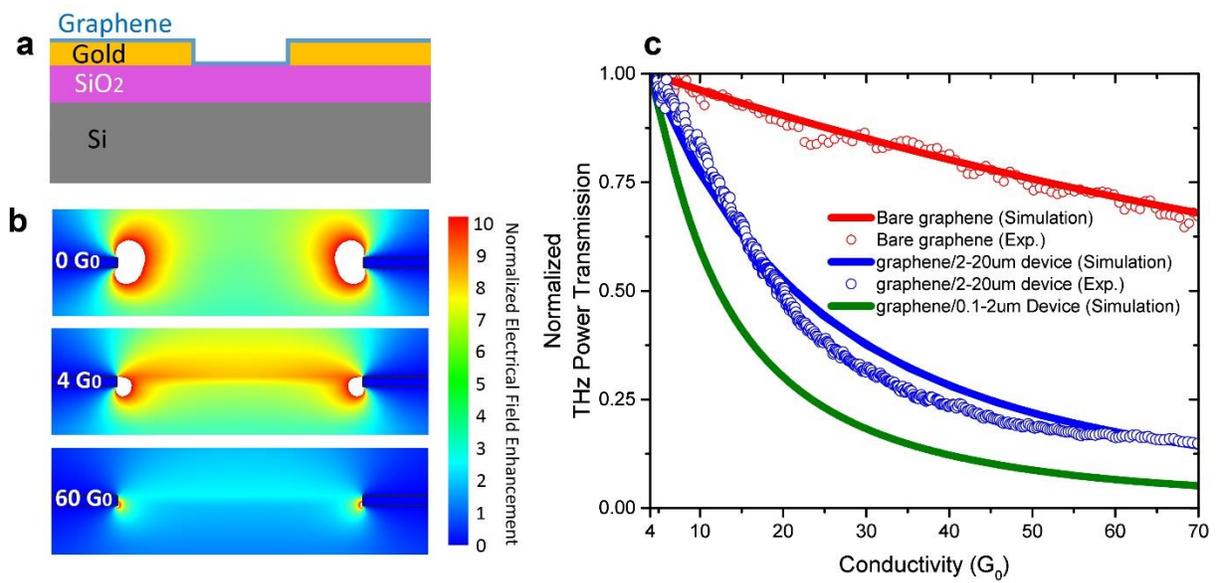

TABLE OF CONTENTS GRAPHIC:

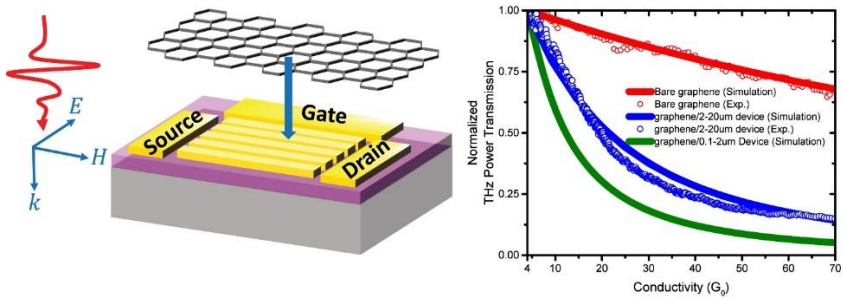